\documentclass[showpacs,preprintnumbers,twocolumn]{revtex4}
\usepackage{amsmath,amssymb,pstricks,wasysym,stmaryrd,enumerate,graphicx}
\usepackage{graphicx}
\begin{document}

\title{From Incoherence to Synchronicity in the Network Kuramoto model}
\author{Alexander C. Kalloniatis\footnote{alexander.kalloniatis@dsto.defence.gov.au}}
\address{
Defence Science and Technology Organisation,
Canberra, Australian Capital Territory, 2600, Australia}

\date{\today}
\begin{abstract}
We study the synchronisation properties of the Kuramoto model of coupled phase oscillators 
on a general network. Here we distinguish the ability of such a system to self-synchronise
from the stability of this behaviour. While self-synchronisation is a consequence of genuine 
non-perturbative dynamics, the stability in dynamical systems is usually accessible by
fluctuations about a fixed point, here taken to be the synchronised solution.
We examine this problem in terms of modes of the graph Laplacian, by
which the absolute Lyapunov stability of the synchronised fixed point is readily demonstrated.
Departures from stability are seen to arise at the next order in fluctuations where
the dynamical equations resemble those for species population models, the
logistic and Lotka-Volterra equations. Methods from these
systems are exploited to analytically derive new critical couplings signalling deviation
from classical stability. We observe in some cases 
an intermediate regime of behaviour, between incoherence
and synchronisation, where system wide periodic behaviours are exhibited.
We discuss these results in light of simulations.
\end{abstract}
\pacs{89.75.Fb,05.45.Xt,89.75.Hc}
\maketitle

\section{Introduction}
\label{intro}
The study of the ability of systems to achieve self-synchronisation -- coherent
behaviour purely through the mutual
interactions of many components -- is a major theme in complex systems 
science. The phenomenon arises in many contexts, from ecology, biology, physics
to social systems. In our case the understanding of self-synchronisation 
through complex networks in
command and control organisations is a major goal \cite{Kall08}.
Notable early work on collective synchronisation includes that of
Wiener \cite{Wie58} and Winfree \cite{Win67}. Kuramoto's elegantly
simple model \cite{Kur84}, upon which we focus, encapsulates much of
the richness expected in self-synchronising systems. Strogatz \cite{Stro00}
gives a short elegant introduction while Acebr{\'o}n {\it et al.} \cite{Ace05}
provide a more thorough review.

The Kuramoto model involves one-dimensional phase
oscillators described by angles $\theta_i$ coupled on a network described by an 
undirected graph $G$ of size $N$ and is given by the first order 
differential equation
\begin{equation}
\dot{\theta}_i = \omega_i - \frac{K}{N} \sum_j A_{ij}
\sin(\theta_i-\theta_j).
\label{KurMod}
\end{equation}
Here $A_{ij}$ is the adjacency matrix encoding the graph structure, $\omega_i$ 
are intrinsic frequencies
of the oscillators, usually drawn from some statistical distribution and $K$ 
is a uniform coupling controlling the degree to 
which adjacent oscillators must mutually adjust with respect to 
each other. At zero coupling the system obviously behaves
incoherently with each oscillator moving in 
simple harmonic motion according to its intrinsic frequency. 
Increasing the coupling enables the population to eventually 
break into synchrony, with some of the
oscillators locking into a single core that 
moves with harmonic motion according to the frequency average over the whole oscillator
ensemble. For the case of an infinite, complete graph and
symmetric and unimodal distribution of intrinsic frequencies Kuramoto was
able to derive an exact analytic 
expression for the critical coupling $K_c$ at which synchrony breaks out
\begin{equation}
K_c={{2}\over{\pi g(0)}}
\label{KurEx}
\end{equation}
where $g(0)$ is the central value of the distribution $g$ from which
the intrinsic frequencies of the individual oscillators are selected.

On a general network no equivalent analytic expression for Kuramoto's
critical coupling has been derived. This paper shall take some steps in that direction.
While analytic methods have been applied to mean-field versions of the model,
such as in \cite{Ich04}, numerical simulations have delivered insights into the pattern
of synchronisation for the complete form of the Kuramoto model on finite general
networks. For random graphs (such as
Erd\"os-Renyi) it is observed by \cite{Gomez07} that, as the
coupling is increased, synchrony first develops locally on separate
hubs which finally merge into a single 
locked system after the coupling has crossed a final threshold. The same authors
find that for ``scale-free'' networks \cite{AlbBar02} 
the synchrony develops at one point in the graph and then
extends to the whole graph as coupling 
is increased. For the Watts-Strogatz ``small world'' 
network \cite{WattsStrog98} synchronisation occurs 
even for small rewiring probability \cite{Hong02}. Such work already shows that
quite specific graph substructures play a subtle 
role in the transient towards synchronisation, both in manifesting local
synchronisation and, possibly, 
stimulating it in other parts of the network. Further insight is
offered by Arenas {\it et al.} \cite{Aren06} who find evidence that the
relevant sub-structures are the modes of the graph Laplacian, an object that
is ubiquitous in network coupled dynamical systems \cite{Chung97,Boll98}.
Indeed, for graphs with hierarchical community structure the order in which
substructures settle into synchronisation is according to the Laplacian eigenvalue,
with the largest mode synchronising first and the smallest stabilising last
\cite{Aren06}.
A more thorough review of numerical studies is provided by \cite{Dor08}.

At the heart of it,
synchronisation and its stability are two separate phenomena in such systems.
These aspects are often confused in the literature, so that, in some cases, stability
criteria (derived perturbatively) are treated as synonymous with synchronisability
(a deeply non-perturbative phenomenon). Intuitively,
the distinction can be appreciated from the perspective of
a potential energy landscape for the entire system.
The evolution of the system from some random initial
condition to complete synchronisation can be visualised as a walk
through this landscape, from the ``edges'' (incoherent behaviour) 
to the ``centre'' (coherence) through a typically rough intermediate
landscape. Synchronisation is the means by which the system passes
through the bulk of the rough landscape close to the centre. Stability
is a local property of the vicinity of the centre.
In the vicinity of the synchronised solution, for stability, 
perturbative methods are applicable.
This is where this paper will focus and provide new analytical insights.
The mechanism for the system finding its way through
the landscape is a genuine nonperturbative problem for which analytic methods
are lacking in the finite general network case though numerical
studies have provided significant insights. This paper will offer an
hypothesis, building on the insights from the existing numerical studies. 

While most stability studies are performed for the infinite network case using the formalism of 
probability densities \cite{Mir05}, in the finite case there is little that has been published
specifically for the Kuramoto model.
We suspect that this is because many researchers may have implicitly appreciated that the 
synchronised solution is Lyapunov stable and decided that there
was little more to be said. We shall 
review this aspect in terms of eigenmodes of the graph Laplacian.
In these terms, the synchronised solution turns out to be just the zero mode of the 
Laplacian while the normal mode fluctuations, to first order, exponentially decay to
zero consistent with the Lyapunov stability criterion. 
More generally, in this framework, we make the observation that
the synchronisation problem can be reframed as a
population competition dynamics. The question then becomes: how do the normal mode interactions
lead to an ``extinction'' of these modes? 
We answer this question by extending 
considerations to second order in the lowest lying Laplacian normal mode fluctuations
and extracting equations more commonly seen in species-population models.
Though we argue that {\it all the low-lying Laplacian normal modes} are significant in the
generation of instabilities in the system, we shall study analytically the
role, respectively, of one and two of the lowest-lying Laplacian normal modes 
in generating instabilities of the synchronised fixed point. The lowest Laplacian
normal mode, whose relationship to measuring connectivity in a network
was noted by Fiedler \cite{Fied73}, has often been argued to be relevant
to the problem, as we shall review below.
In our analysis we show that the dynamics of a single lowest
mode are governed by a modification of the {\it logistic} equation
while for two lowest modes a form of {\it Lotka-Volterra} equations are obtained. 
Though, for many complex networks, these are rather artificial cases,
we derive critical couplings here to illustrate the types of dynamics that can occur.
Indeed, we identify in some cases a previously (to our knowledge) unnoticed
regime lying between incoherence and coherence, a type of ``edge-of-chaos''
patterned dynamics. We provide support for this by showing examples of simulations
for this system.

The paper is structured as follows.
In the next section we recast the Kuramoto model in terms of modes of the
graph Laplacian and review the Lyapunov stability of the system.
Then we develop the problem to second order in fluctuations and
study several cases for the structure of the lowest normal modes of the Laplacian.
We derive a number of critical couplings for several special cases. 
We subsequently analyse the behaviour of an order parameter
for synchronisation of the system in these terms and discuss this in
light of simulations. The final section will discuss our insights into
the general mechanism for synchronisation and the subtleties of generalising
these results to the complete infinite graph, originally studied by Kuramoto.

\section{Laplacian eigenmodes and stability}
\label{Lapmodes}

\subsection{Comparison of Kuramoto model with other coupled dynamical models}
Let us begin by contrasting the Kuramoto model, Eq.(\ref{KurMod}), with another
class of coupled dynamical system which has been studied intensively for some time:
\begin{eqnarray}
\dot{\phi_i} = f(\phi_i) + \sigma \sum_j A_{ij} \left[
  g(\phi_j)-g(\phi_i)\right].
\nonumber
\end{eqnarray}
We shall consistently use the graph (or combinatorial) Laplacian \cite{Chung97,Boll98},
\begin{equation}
L_{ij} = d_i \delta_{ij} - A_{ij},
\label{LaplDef}
\end{equation}
with $d_i$ is the degree of node $i$.
The spectrum of the Laplacian is positive semi-definite, so that
eigenvalues can be ordered from smallest to largest:
\begin{equation}
0=\lambda_0 \leq \lambda_1 \leq \dots \leq \lambda_{N-1}.
\label{evalhier}
\end{equation}
(In some of the literature the numbering of modes
starts at one and not zero.)
In these terms, the dynamical system can be rewritten
\begin{equation}
\dot{\phi_i} = f(\phi_i) - \sigma \sum_j L_{ij} g(\phi_j).
\label{DynSys2}
\end{equation}
Synchronisation is represented by fixed point solutions $\phi_i=s$ of Eq.(\ref{DynSys2}). 
Fluctuations $\eta_i$ about $s$ satisfy then
\begin{eqnarray}
\dot{\eta}_i = f'(s)\eta_i - \sigma g'(s) \sum_j  L_{ij} \eta_j.
\nonumber
\end{eqnarray}
Going to a basis of eigenmodes for the fluctuations leads to a criterion for
stability of the synchronised solution:
\begin{eqnarray}
\int_0^t d\tau \left[
f'(s(\tau)) - \sigma g'(s(\tau)) \lambda_r \right] < 0
\nonumber
\end{eqnarray}
for all Laplacian modes $r$.
This ``master stability equation'' was derived by
Pecora and Carroll \cite{Pec98}. Many others have exploited this approach 
to study a variety of networks, for example \cite{various}. 
Significantly, we see that, depending on the network and the nature of the
functions $f$ and $g$, there may be directions in Laplacian mode space for which
this condition is not satisfied. Therefore, within a first order perturbation calculation
we can determine a condition for which synchronisation can exist and is stable.
It is tempting on this basis to see synchronisation purely through the lens of Lyapunov stability.

Unfortunately, the Kuramoto model Eq.(\ref{KurMod}) is very different from this 
system. This is because the interaction in the Kuramoto model is a
{\it function of differences} rather than a {\it difference of
functions}. Practically, it means that the dynamics are not automatically
diagonalised in the Laplacian eigenmode basis.
However, small fluctuations about 
the synchronisation point can be diagonalised in terms of the
Laplacian. 

\subsection{Representation in Laplacian Basis}
Before diving into this, let us present some more details on the Laplacian.
The degeneracy of zero eigenmodes of the Laplacian 
corresponds to the number of
disconnected components of the graph. We shall concern ourselves with
connected graphs, so there will be a single zero mode. 
The corresponding zero mode eigenvector, within
a system of orthonormal eigenvectors, is 
\begin{eqnarray}
\nu^{(0)T} &=& \frac{1}{\sqrt{N}} (1,1,\dots,1) \label{zerovec}, \\
L \nu^{(r)} &=& \lambda_r \nu^{(r)}, \nonumber \\
\nu^{(r)} \cdot \nu^{(s)}& =& \delta_{rs} . \nonumber
\end{eqnarray}
The first nonzero eigenmode, here corresponding to $\lambda_1$, 
is called the {\it Fiedler} with the 
eigenvalue itself known as the algebraic connectivity \cite{Fied73}. 
The eigenmode itself is used in other contexts as a measure of network
fragility \cite{Ding01}.
Though we are dealing with unoriented graphs, it is convenient to rewrite the Laplacian  
in terms of a signed incidence matrix $B$ as follows. 
Let links, which may be assigned an arbitrary direction for this purpose, be 
labelled by the index $a$. Matrix elements of $B$ are $B_{ia}$ where
\begin{eqnarray}
B_{ia} = +1, 0, -1 \nonumber
\end{eqnarray}
according to whether link $a$ is incoming to, not connected with or
outgoing from node $i$. The Laplacian is then given by
\begin{eqnarray}
L=BB^T. \nonumber
\end{eqnarray}
The signed incidence matrix has deep mathematical properties, but one
role that it plays is to transform between objects respectively in the
node and link spaces, including the eigenvectors:
\begin{eqnarray}
e^{(r)} &=& B^T \nu^{(r)} \\
e^{(r)}\cdot e^{(s)} &=& \lambda_r \delta_{rs}. \nonumber
\end{eqnarray}
In terms of the original eigenvectors 
we can expand a phase angle of the Kuramoto model in the Laplacian
basis
\begin{eqnarray}
\theta_i(t) = \sum_{r=0} \alpha_r(t) \nu^{(r)}_i. \nonumber
\end{eqnarray}
Inserting this in the equation of motion Eq.(\ref{KurMod}) we arrive
at equations for the components:
\begin{eqnarray}
\alpha_0 &=& \sqrt{N} \bar{\omega} t \label{zeromode} \\
\dot{\alpha_r}(t) &=& \omega^{(r)} - \sigma \sum_{a} e^{(r)}_a 
\sin\left( \sum_{s>0} e^{(s)}_a \alpha_s(t) \right); \nonumber \\
&& r\neq0
\label{KurLapl}
\end{eqnarray}
where from this point we summarise the coupling as $\sigma=K/N$. Also
implicit in Eq.(\ref{KurLapl}) is that we treat the $\omega_i$ as an
$N$-dimensional vector, with $\bar\omega$ the average over the distribution and
\begin{eqnarray}
\omega^{(r)}=\omega \cdot \nu^{(r)}
\nonumber
\end{eqnarray}
represents the projection of the vector of frequencies $\omega_i$ onto the $r$-th
Laplacian mode.

\subsection{Interpretation of normal mode dynamics}
From Eq.(\ref{zeromode}), we see that the zero mode is just the synchronised mode
and that it {\it completely decouples from the normal modes}. Therefore,
in terms of Laplacian modes there is always a component moving harmonically with
the average frequency. Posed this way, the concern ceases to be how synchronisation is
generated but how the normal mode dynamics in Eq.(\ref{KurLapl}) are suppressed. 
Indeed, we can reformulate the problem in the language of another classical complex systems
problem: that of inter/intra-species population dynamics. Recall that in this context, 
populations can grow, die out and compete with others in predator-prey relationships.
Within such interactions, oscillating population numbers for co-existing species can occur. 

In this language, then, pure oscillator synchronisation is identical to 
normal {\it eigenmode population extinction} and 
complete oscillator incoherence is identical to multiple species 
{\it normal eigenmode population co-existence}.
Later in the paper, we shall refine this relationship further and identify specific
cases of behaviours according to interacting population dynamics 
manifested here in the competing Laplacian normal eigenmodes.

\subsection{Lyapunov Stability}
We can now review standard stability arguments. Let the system be close to complete
synchronisation, now characterised by the zero mode,
\begin{eqnarray}
\theta_i(t) \approx \alpha_0(t) \nu_i^{(0)} + {\rm{small}},
\nonumber
\end{eqnarray}
with small fluctuations
being some superposition of the normal modes. Expanding the 
interaction term for small normal modes to leading order, we obtain
\begin{eqnarray}
\dot{\alpha_r}(t) = \omega^{(r)} - \sigma \lambda_r
\alpha_r(t); r\neq 0. 
\label{firstapproxeq}
\end{eqnarray}
The solution to this is just
\begin{equation}
\alpha_r(t) = {{\omega^{(r)}}\over{\sigma \lambda_r}}
(1-e^{-\sigma\lambda_r t}) +
\alpha_r(0) e^{-\sigma \lambda_r t}.
\label{firstapprox}
\end{equation}
Part of this equation was obtained by Arenas {\it et al.} \cite{Aren06}, who
only presented the second term as their focus was on the hierarchy of
suppression of modes. This supports the idea that the sequence of
clusters achieving synchronisation occurs sequentially in time through the
hierarchy of eigenvalues in Eq.(\ref{evalhier}). Specifically,
fluctuations in the highest 
mode of the Laplacian, corresponding to $\lambda_{N-1}$, 
are suppressed most quickly; the last mode to be
suppressed is the 
lowest non-zero mode, the Fiedler with eigenvalue $\lambda_1$. 
{\it If} suppression is achieved, the modes become constant
\begin{equation}
\alpha_r^{(\infty)} = {{\omega^{(r)}}\over{\sigma \lambda_r}}.
\label{alphaconst}
\end{equation}
Note that the complete graph has a completely degenerate Laplacian normal mode
spectrum: $\lambda_i=N, \ i=1, \dots, N-1$. The implications of this
degeneracy are discussed in the final section. We observe in Eq.(\ref{alphaconst}) that
the regime of applicability of synchronisation is characterised by small $\alpha_r^{(\infty)}$,
namely
\begin{equation}
\sigma > {{\omega^{(r)}}\over{\lambda_r}}
\label{1stcritcoupl}
\end{equation}
which can be seen as a (very) weak bound on critical values of the coupling such that synchronisation
can occur. However, this bound does not indicate what types of transitions can occur.

We see that in the vicinity of the synchronized solution the system is
Lyapunov stable in all directions in Laplacian mode space: the Lyapunov exponents
are exactly the Laplacian eigenvalues multiplied by the coupling so that there are
no regimes of coupling or structure whereby they may change sign.
This has also been observed by Jadbabaie {\it et al.} \cite{Jad04} using different methods.
These authors also derive certain bounds on the coupling, related to
our bound Eq.(\ref{1stcritcoupl}), which, however, do not readily return
the special case derived by Kuramoto for the infinite complete graph. We
suspect this is a consequence of the absolute Lyapunov stability of the system
and return to this at the end of the next section.

\section{Second order approximations}
\label{nextorder}

In the following we extend the above approximation to next order in the Taylor expansion 
of the interaction in the equations of motion and
making certain simplifying assumptions on the structure of the lowest modes of the
Laplacian. Specifically, we expand the sine in Eq.(\ref{KurLapl}) out to the cubic term,
$\sin x=x-x^3/6$. This means that our approximations are valid for values of $\alpha_r$
larger than considered in standard stability analysis thus far,
out to values of $x\sim \pi/2>1$.
Furthermore we shall assume sufficient time to have elapsed in the running of
the system dynamics so that {\it most} Laplacian modes
are suppressed to $\alpha_r^{(\infty)}$. The exception to this will be
a number of the lowest modes which will be taken to be dynamical. 
In other words, we exploit the observation in numerical simulations that 
modes with small Laplacian eigenvalue synchronise later, though we make no assumptions
about the order of synchronisation of the higher modes.
We shall now explore the impact of these few small eigenvalue dynamical modes on stability.

\subsection{One dynamical mode}
\label{onemode}
\subsubsection{Exact solutions}
In this case we have a single mode that remains time-dependent, namely that corresponding to
the Fiedler, $\alpha_1(t)$. All other modes $r>1$ assume their
steady-state values Eq.(\ref{alphaconst}). We assume in this part that
there is no degeneracy for $\lambda_1$, which will be relaxed in the next section.
Thus we separate the
mode sum in Eq.(\ref{KurLapl}) into two parts,
\begin{equation}
\sum_{s>0} e^{(s)}_a \alpha_s(t) = 
e^{(1)}_a \alpha_1(t)
+ \sum_{s>1} e^{(s)}_a \alpha_s^{(\infty)} ,
\label{modesep}
\end{equation}
and insert this into Eq.(\ref{KurLapl}). Within the expansion of the sine 
we drop all terms cubic in any given mode 
and those quadratic in the
constant modes $\alpha_s^{(\infty)}$. To this order then we extract the equation
\begin{equation}
\dot{\alpha_1}(t) = \omega^{(1)} - \sigma \lambda_1 \alpha_1(t) + 
\sigma x_1 \alpha_1(t)^2,
\label{nextapprox}
\end{equation}
where we define constants summarising the dynamical and structural content of the
system
\begin{eqnarray}
x_r &=& \sum_{s>1} {c_{rs}{\omega^{(s)}\over{2\sigma \lambda_s}}}, \nonumber
\\
c_{rs} &=& \sum_a (e^{(r)}_a)^3 e^{(s)}_a, \ s\neq r. \nonumber
\end{eqnarray}
Eq.(\ref{nextapprox}) 
is a variation on the well-known logistic equation. 
We further introduce the notation (the first being the discriminant of a quadratic equation):
\begin{eqnarray}
\Delta_1 &=& \sigma^2 \lambda_1^2 - 4 \sigma \omega^{(1)}x_1, \nonumber \\
P_1(0) &=& { { 2 \sigma x_1 \alpha_1 (0) - (\sigma\lambda_1+\sqrt{\Delta_1}) }\over 
{2 \sigma x_1 \alpha_1(0) - 
(\sigma\lambda_1-\sqrt{\Delta_1})}}.
\nonumber
\end{eqnarray}
The solution to Eq.(\ref{nextapprox}) depends on the sign of the discriminant $\Delta_1$. 
When the coupling strength $\sigma$ is varied, say, from large to down
small values the discriminant changes sign, from
positive, through zero, zero to negative values. 

For large $\sigma\lambda_1$ the solution is
\begin{equation}
\alpha_1(t) = {{(\sigma\lambda_1+\sqrt{\Delta_1}) - 
(\sigma\lambda_1-\sqrt{\Delta_1})
 P_1(0) e^{\sqrt{\Delta_1}t}}\over
  {2 \sigma x_1(1- P_1(0) e^{\sqrt{\Delta_1}t})}}.
\label{alphstable}
\end{equation}
This will return the first order solution when $\lambda_1 \ll \lambda_2$. 
Then $x_1\approx 0$ and the right hand side of Eq.(\ref{alphstable}) can be
expanded for small $x_1$ giving Eq.(\ref{firstapprox}).
This reflects the behaviour that the solution to Eq.(\ref{nextapprox}) 
starts from the initial
condition and then is exponentially suppressed to the constant value
Eq.(\ref{alphaconst}). 
There can occur points where, at some fixed
time, the denominator of Eq.(\ref{alphstable}) vanishes; 
the solution shoots to negative
infinity and then returns to stabilise at the constant value
Eq.(\ref{alphaconst}). 

When the discriminant is negative the solution takes the form
\begin{eqnarray}
&&\alpha_1(t) = \frac{1}{2\sigma x_1} \times \nonumber \\
&& \left[ \sigma \lambda_1 + \tan\left(\arctan(2 \sigma x_1 \alpha_1(0)
  - \sigma\lambda_1) + \sqrt{-\Delta}_1 \frac{t}{2}\right) \right]. \nonumber \\
\label{alphunstable}
\end{eqnarray}
This displays both ``accidental'' divergent behaviour, when the argument of the
tangent becomes  $\pi/2$ and, more significantly, periodicity in time. 
In this regime then, the normal modes
are dominated by a limit cycle and the lowest Laplacian mode {\it never converges to
steady state}.

Therefore the condition $\Delta_1=0$ separates two distinct
regimes of behaviour: for $\Delta_1<0$ complete synchronisation is not attained while for 
$\Delta_1>0$ convergence to the fixed point occurs. We can thus extract a critical coupling:
\begin{equation}
\sigma_c = 2 \sum_{s>1} { { \omega^{(1)} c_{1s}
    \omega^{(s)} } \over {\lambda_1^2 \lambda_s}}.
\label{newcrit}
\end{equation}
This reflects a genuine point at which the 
dynamics changes {\it distinctly} in character in contrast to the bound Eq.(\ref{1stcritcoupl}) 
derived purely within a first order stability analysis. 

\subsubsection{Equilibrium analysis}
\label{subsectEquil}
Even though we have analytical solutions here it is worth reinforcing
these results by looking at the equilibrium dynamics in this regime.
Setting $\dot{\alpha_1}=0$ in Eq.(\ref{nextapprox}), two solutions result:
\begin{eqnarray}
\alpha_1^{\pm} = {{\sigma\lambda_1 \mp \sqrt{\Delta_1}} 
\over {2\sigma x_1}}.
\nonumber
\end{eqnarray}
In fact these match on to the analytical solution for $t\rightarrow \pm\infty$ of 
Eq.(\ref{alphstable}):
\begin{eqnarray}
\alpha_1^{\pm}=\lim_{t\rightarrow \pm\infty} \alpha_1(t).
\nonumber
\end{eqnarray}
As a preview to the next section, let us examine stability about these equilibria.
Inserting
\begin{eqnarray}
\alpha_1 = \alpha_1^{\pm} + \chi^{\pm}
\nonumber
\end{eqnarray}
into Eq.(\ref{nextapprox}) we obtain the order $\chi$ equation
\begin{eqnarray}
\dot{\chi} &=& (2 \sigma x_1 \alpha^{\pm} - \sigma\lambda_1) \chi^{\pm} \nonumber\\
&=& \mp \sqrt{\Delta_1} \chi^{\pm}
\nonumber
\end{eqnarray}
The solution manifests a Lyapunov-like exponent:
\begin{eqnarray}
\chi^\pm = \chi^\pm (0) e^{\mp \sqrt{\Delta_1} t}
\nonumber
\end{eqnarray}
Again, we are only interested in the $t\rightarrow \infty$ behaviour, namely in $\chi^+$.
Examining the strong coupling case $\sigma> 1/\lambda_1$ we obtain
\begin{eqnarray}
\alpha_1^{+} &\rightarrow& {{\omega^{(1)}}\over {\sigma \lambda_1}} \label{1mode+limit} \\
\alpha_1^{-} &\rightarrow& {{\lambda_1}\over {x_1}}- {{\omega^{(1)}}\over {\sigma \lambda_1}}.
\label{1mode-limit}
\end{eqnarray}
Thus, for $t\rightarrow\infty$, Eq.(\ref{1mode+limit}) coincides with the original lowest order
steady state result, Eq.(\ref{alphaconst}). 

We now appreciate the two regimes of behaviour seen in
Eqs.(\ref{alphstable},\ref{alphunstable}) in another light: $\Delta_1>0$
manifests stability, namely convergence as $t\rightarrow \infty$ 
to the equilibrium fixed point while $\Delta_1<0$ 
gives an imaginary phase, namely a limit cycle.

\subsection{Two dynamical modes}
Let us consider initially now the lowest two normal modes, $\alpha_1$ and $\alpha_2$,
to be dynamical with all others having reached steady state. Initially we consider all terms
linear in the constant modes $\alpha_r^{(\infty)}$ but up to quadratic in the dynamical modes
$\alpha_s(t)$ ($s=1,2$). The full system to this order has the form
\begin{eqnarray}
\dot{\alpha_1}(t) &=& \omega^{(1)} - \sigma \lambda_1 \alpha_1(t) + 
\sigma \alpha_1^2 X_1 + \sigma \alpha_1(t) \alpha_2(t) Y_1 \nonumber \\
&& + \sigma \alpha_2(t)^2 Z_1 \nonumber \\
\dot{\alpha_2}(t) &=& \omega^{(2)} - \sigma \lambda_2 \alpha_1(t) + 
\sigma \alpha_2^2 X_2 + \sigma \alpha_1(t) \alpha_2(t) Y_2 \nonumber \\
&& + \sigma \alpha_1(t)^2 Z_2,
\label{2dfull}
\end{eqnarray}
with a set of dynamical and structural constants
\begin{eqnarray}
X_r & \equiv & \sum_{s>2} C_{rs} {{\omega^{(s)}\over{2\sigma \lambda_s}}} \nonumber \\
Y_r & \equiv & \sum_{s>2} D_{rs} {{\omega^{(s)}\over{\sigma \lambda_s}}} \nonumber \\
Z_r & \equiv &  \sum_{s>2} E_{rs} {{\omega^{(s)}\over{2\sigma \lambda_s}}} \nonumber \\
C_{rs} & \equiv & \sum_a (e^{(r)}_a)^3 e^{(s)}_a \nonumber \\
D_{1s} & \equiv & \sum_a (e^{(1)}_a)^2 e^{(2)}_a e^{(s)}_a \nonumber \\
E_{1s} & \equiv & \sum_a e^{(1)}_a (e^{(2)}_a)^2 e^{(s)}_a \nonumber \\
D_{2s} &=& E_{1s} \nonumber \\
E_{2s} &=& D_{1s} \nonumber
\end{eqnarray}
by analogy with $x_r$ and $c_{rs}$ for the one-mode case. 

Eqs.(\ref{2dfull}) are not amenable to analytic solution.
However, analogously to the analysis of Sect.\ref{subsectEquil}, the equilibria and their stability
are open to analysis (as is standard in predator-prey models). In other words, the intention is
to study the
case where $\dot{\alpha_r}=0$. Denoting such solutions to Eqs.(\ref{2dfull}) by $\alpha_r^{(u)}$,
where $u$ represents some label for the different possible solutions, 
we then examine fluctations $\chi^{(u)}_r$ about these, namely
solutions of the form $\alpha_r=\alpha_r^{(u)}+\chi_r^{(u)}$ keeping only terms of order $\chi$.
The system Eqs.(\ref{2dfull}) with LHS set to zero is soluble 
however the solutions are extremely lengthy and cataloging the type of behaviours that can
occur is difficult, though we should expect the full range of dynamical possibilities:
stability, limit cycles, tori and instability. 
To illustrate this we consider a further truncation of the system, where only
cross-interactions are kept. We comment later on the possible generality of our insights from this
analysis.

\subsubsection{Cross-interactions only, non-degenerate case}
Treating the case that $\lambda_1 \neq \lambda_2$, in the first instance, 
we are down to the coupled equations:
\begin{eqnarray}
\dot{\alpha_1}(t) &=& \omega^{(1)} - \sigma \lambda_1 \alpha_1(t) + 
\sigma \alpha_1(t) \alpha_2(t) Y_1 \nonumber \\
\dot{\alpha_2}(t) &=& \omega^{(2)} - \sigma \lambda_2 \alpha_1(t) + 
\sigma \alpha_1(t) \alpha_2(t) Y_2 ,
\label{2dapprox}
\end{eqnarray}

Apart from the first term, we recognise in
Eqs.(\ref{2dapprox}) the two dimensional Lotka-Volterra system for predator-prey
dynamics \cite{Kap95}. That we have arrived at the next step in generalisation of population
models after the logistic equation is consistent with our picture of the normal mode dynamics.

The solutions of Eqs.(\ref{2dapprox}) with  $\dot{\alpha}_1=\dot{\alpha}_2=0$ 
are compactly represented by introducing a new discriminant:
\begin{eqnarray}
\Delta_2 = (\sigma \lambda_1 \lambda_2 - (\omega^{(1)} Y_2 - \omega^{(2)} Y_1))^2
- 4 \sigma \lambda_1 \lambda_2 \omega^{(2)} Y_1.
\nonumber
\end{eqnarray}
Eqs.(\ref{2dapprox}) exhibit two equilibria:
\begin{eqnarray}
\alpha_1^{(\pm)} &=& { { \sigma \lambda_1 \lambda_2 + (\omega^{(1)}Y_2 - \omega^{(2)} Y_1) 
\mp \sqrt{\Delta_2}}
\over {2 \sigma \lambda_1 X_2}} \nonumber \\
\alpha_2^{(\pm)} &=& { { \sigma \lambda_1 \lambda_2 - (\omega^{(1)}Y_2 - \omega^{(2)} Y_1) 
\mp \sqrt{\Delta_2}}
\over {2 \sigma \lambda_2 Y_1}}.
\label{2dequil}
\end{eqnarray}

We consider now the fluctuations $\chi_r^{(u)}$ about these solutions by
insertion into Eq.(\ref{2dapprox}). We arrive at a fluctuation matrix whose 
eigenvalues give Lyapunov exponents for the equilibria Eqs.(\ref{2dequil})
which have the form
\begin{eqnarray}
l_1^{(\pm)} &=& - {1\over{4\lambda_1 \lambda_2}}
\left( \sqrt{(A\pm B)^2 \mp C} \pm (A\pm B) \right) \nonumber \\
l_2^{(\pm)} &=& + {1\over{4\lambda_1 \lambda_2}}
\left( \sqrt{(A \pm B)^2 \mp C} \mp (A \pm B) \right) 
\label{lyapunovs}
\end{eqnarray}
with
\begin{eqnarray}
A &=& (\lambda_1+\lambda_2)\sqrt{\Delta_2} \nonumber \\
B &=& \sigma \lambda_1 \lambda_2 (\lambda_1+\lambda_2) \nonumber \\
&& - (\lambda_2-\lambda_1) (\omega^{(1)}Y_2-\omega^{(2)} Y_1) \\
C &=& 16 \sigma \lambda_1^2\lambda_2^2 \sqrt{\Delta_2}. \nonumber
\end{eqnarray}

To get a handle on the equilibria in the absence of explicit analytic solutions it is worth examining 
first the strong coupling limit $\sigma>1/\lambda_r$. For the equilibria we obtain from
Eqs.(\ref{2dequil}) 
\begin{eqnarray}
\alpha_1^{(+)} &\rightarrow& {{\omega^{(1)}}\over {\sigma\lambda_1}} \nonumber \\
\alpha_2^{(+)} &\rightarrow& {{\omega^{(2)}}\over {\sigma\lambda_2}} \nonumber \\
\alpha_1^{(-)} &\rightarrow& {{\lambda_2}\over{Y_2}}- {{\omega^{(2)}Y_1}\over{\sigma\lambda_1Y_2}} 
\nonumber \\
\alpha_2^{(-)} &\rightarrow&  {{\lambda_1}\over{Y_1}}- {{\omega^{(1)}Y_2}\over{\sigma\lambda_2Y_1}}
\nonumber 
\end{eqnarray}
while the Lyapunov exponents Eqs.(\ref{lyapunovs}) give:
\begin{eqnarray}
l_1^{(+)} & \rightarrow& - \sigma \lambda_1 \nonumber \\
l_2^{(+)} & \rightarrow& - \sigma \lambda_2 \nonumber \\
l_1^{(-)} & \rightarrow& - \sigma \sqrt{\lambda_1\lambda_2} \nonumber \\
l_2^{(-)} & \rightarrow& + \sigma \sqrt{\lambda_1 \lambda_2}. \nonumber 
\end{eqnarray}
We recognise then that $\alpha_r^{(+)}$ converge to the $t\rightarrow\infty$ form of the
lowest order solutions. Our concern being with $t>0$ we therefore discard the solutions
$\alpha_r^{(-)}$.

Noting that $A,C\geq 0$, we test the possible signs of the exponents in Eqs.(\ref{lyapunovs}). 
Consider, firstly,
the case of the discriminant $\Delta_2>0$. Then it is straightforward to derive the following
cases:
\begin{eqnarray}
A+B>0 & \Rightarrow & l_1^{(+)} < 0, \ l_2^{(+)} < 0 \nonumber \\
A+B<0 & \Rightarrow & l_1^{(+)} > 0 {\rm {\ or \ complex}}, \ l_2^{(+)} > 0 {\rm {\ or \ complex}} 
\nonumber .
\end{eqnarray}
The complex values occur when $(A+B)^2<C$. Therefore, for $\Delta_2>0$, 
$\alpha_1$ and $\alpha_2$ will either be {\it both} stable or both
display limit cycles. In the latter case, there can be real parts with positive sign
in $l_2^{(+)}$ of Eqs.(\ref{lyapunovs}),
indicating instability. We shall illustrate this in the simpler degenerate case below.

More interestingly, there is a precise coincidence between the ``critical point'', given by
$\Delta_2=0$, and an ``edge-of-chaos'' regime defined by vanishing Lyapunov
exponents, $l_2^{(1)}=l_2^{(2)}=0$. 
The boundaries of this regime are governed by the two solutions to $\Delta_2=0$:
\begin{eqnarray}
\sigma_{\pm} = {1 \over{\lambda_1\lambda_2}} 
(\omega^{(1)}Y_2 + \omega^{(2)}Y_1 \pm 2\sqrt{\omega^{(1)}\omega^{(2)}Y_1Y_2}).
\nonumber 
\end{eqnarray}
Evidently $\sigma_- < \sigma_+$ and the two values indicate sign changes in $\Delta_2$:
positive for $\sigma < \sigma_-$ and $\sigma > \sigma_+$ and 
negative for $\sigma_- < \sigma < \sigma_+$. Noting that at $\sigma=0$ we
have $\Delta_2=(\omega^{(1)}Y_2-\omega^{(2)}Y_1)^2>0$ then either both critical points,
$\sigma_{\pm}$, are positive or both negative (and therefore
unphysical as we consider only locally attractive interactions): 
if one critical point appears then there will be a second. 

We conclude therefore that between incoherence for $\sigma \approx 0$ and coherence
for $\sigma \gg 1/\lambda_r$ there will be an intermediate regime characterised by
oscillatory behaviour whose frequency is not related to any individual oscillator
but to the collective behaviour of the low-lying graph Laplacian modes.

\subsubsection{Degenerate case}
In the case that two lowest Laplacian eigenvalues are equal, $\lambda_1=\lambda_2$, 
a further simplification occurs. The Lyapunov exponents relevant for $t>0$
turn out to be:
\begin{eqnarray}
l_1^{(+)} &=& -{1\over{2\lambda_1}} 
\left[ |\sqrt{\Delta_2} - \sigma \lambda_1^2| + (\sqrt{\Delta_2} + \sigma \lambda_1^2)\right] 
\nonumber \\
l_2^{(+)} &=& {1\over{2\lambda_1}} 
\left[ |\sqrt{\Delta_2} - \sigma \lambda_1^2| - (\sqrt{\Delta_2} + \sigma \lambda_1^2)\right].  
\nonumber
\end{eqnarray}

Consider firstly the case of $\Delta_2>0$.
For $\sqrt{\Delta_2}>\sigma\lambda_1^2$ we have:
\begin{eqnarray}
l_1^{(+)} &=& - {{\sqrt{\Delta_2}}\over {\lambda_1}}\nonumber \\
l_2^{(+)} &=& - \sigma \lambda_1 . \nonumber
\end{eqnarray}
On the other hand, for $\sqrt{\Delta_2}<\sigma\lambda_1^2$ the exponents swap.

When the discriminant is negative we must take care with the modulus. Writing
\begin{eqnarray}
\sqrt{\Delta_2} = i \sqrt{|\Delta_2|} \nonumber 
\end{eqnarray}
for this case, we have
\begin{eqnarray}
 |\sqrt{\Delta_2} - \sigma \lambda_1^2| &=& |\sigma\lambda_1^2 - i \sqrt{|\Delta_2|}|\nonumber \\
&=& \sqrt{ |\Delta_2| +\sigma^2 \lambda_1^4}. \nonumber
\end{eqnarray}
Thus
\begin{eqnarray}
l_1^{(+)} &=& - {1\over{2\lambda_1}} 
\left[ \sqrt{ |\Delta_2| +\sigma^2 \lambda_1^4} + \sigma\lambda_1^2 + i \sqrt{|\Delta_2|} \right]
\nonumber \\
l_2^{(+)} &=&  {1\over{2\lambda_1}} 
\left[ \sqrt{ |\Delta_2| +\sigma^2 \lambda_1^4} - \sigma\lambda_1^2 - i \sqrt{|\Delta_2|} \right].
\nonumber
\end{eqnarray}
The fixed points are therefore ``hyperbolic'', having non-vanishing real parts.
Indeed, since $\sqrt{ |\Delta_2| +\sigma^2 \lambda_1^4} - \sigma\lambda_1^2 >0$
for non-vanishing discriminant,
the sign of the real part of the second mode will evidently be positive
signalling a genuine instability.

In summary then, for $\Delta_2>0$ we obtain stability, namely convergence to 
the steady state while for $\Delta_2<0$, we obtain limit cycle behaviour
for both modes mixed with instability in
the second mode. Once again, $\Delta_2=0$ separates two distinct regimes of behaviour, giving two
possible critical couplings:
\begin{eqnarray}
\sigma_{\pm} = {1 \over{\lambda_1^2}} 
(\omega^{(1)}Y_2 + \omega^{(2)}Y_1 \pm 2\sqrt{\omega^{(1)}\omega^{(2)}Y_1Y_2}).
\nonumber 
\end{eqnarray}
The above analysis can be reproduced here.

Finding manageable explicit solutions beyond these truncations is difficult. However, at a general
level we can identify the structures which are significant in the results we have seen. Typically
the critical couplings arise, in the cases considered here, from the discriminants 
corresponding to quadratic equations for the equilibria. The equilibria for
the general system Eqs.(\ref{2dfull}) evidently will satisfy higher degree polynomial equations,
for which corresponding discriminants occur. For example, treating the first of Eqs.(\ref{2dfull}) as
an equation for $\alpha_1$ in terms of $\alpha_2$ we have initially a quadratic equation.
Substitution of $\alpha_1$ into the second equation will lead to an equation involving 
square roots of the discriminant which can be eliminated by squaring the equation. This
will generate a quartic with its own discriminant. We thus foresee cases where the
Lypunov exponents for the various equilibria will develop imaginary parts when the 
discriminants are less than zero. What is more difficult to oversee is whether there
are cases when the Lypunov's are purely real and {\it positive}, signals of actual chaotic instability
to this order.

\subsection{More dynamical modes}
Going beyond the case of two dynamical modes at present appears difficult within 
analytical approaches. However, the identification of the dynamics as being
essentially a modification of multi-species competitive Lotka-Volterra equations enables
us to exploit a general result by Smale \cite{Smal76}. Here it is known
that for systems of more than five population species the equilibria can be
of any type: fixed-point, limit cycle, torus or attractors. However, this is
just a statement that anything is possible. Thus, whereas thus far we have
only seen limit cycles prohibit complete synchronisation, for more than two closely
spaced low-lying Laplacian modes we can expect tori and genuine chaos to be
exhibited in the dynamics. The more severe these behaviours, the further
from partial synchronisation and the close to true incoherence the system approaches.
We can also expect subtle dynamical effects, as seen in multi-species Lotka-Volterra systems, such as 
purely interaction generated spatio-temporal patterns (or
spontaneous symmetry breaking) \cite{SWA05}. What we lack at this stage are criteria
that may give insights into the boundaries between different regimes.

Also, our considerations cannot yet be extended to the case of the complete
graph: the above approximations assume a separation between static and dynamic modes
which cannot be reconciled with the total degeneracy of the Laplacian eigenvalue spectrum 
for the complete graph. The insight that the normal mode dynamics can
be mapped onto population models is still valid nevertheless, though the specific 
modified Lotka-Volterra equations
considered above cannot be taken to hold. For that reason it is premature to seek to extrapolate
the above results for critical couplings to check against Kuramoto's analytic result.

\section{Order Parameter}
We now explore the impact of the above solutions on an order parameter characterising 
the possible phase of the system. Jadbabaie {\it et al.} \cite{Jad04} propose a generalisation of
Kuramoto's order parameter for the general network:
\begin{eqnarray}
r^2(t) = {{N^2 - 2E + 2 \sum_a \cos B_{ai}\theta_i(t)}\over{N^2}},
\nonumber
\end{eqnarray}
with $E$ the number of edges in the network.
This can be rendered more transparently by using the identity $\cos x=1 - 2\sin^2 x/2$,
converting to the Laplacian basis and explicitly using the form of the zero mode
eigenvector (which makes it drop out of $r^2$), giving
\begin{equation}
r^2(t)=1 - \frac{4}{N^2} \sum_a \sin^2 \left( \frac12 \sum_{r\neq 0} \alpha_r(t) e_a^{(r)} \right).
\label{r-Lapl}
\end{equation}
We thus see explicitly that complete synchronisation as characterised by complete
extinction of normal modes, $\alpha_r=0$, gives $r^2=1$. However this is
a stronger condition than the steady state behaviour, $\alpha_r^{(\infty)}$, for which
the order parameter is constant
\begin{eqnarray}
r^2\rightarrow (r^{(\infty)})^2 \equiv 1 - \frac{4}{N^2} \sum_a \sin^2 \left( \frac12 \sum_{r\neq 0}
\alpha_r^{(\infty)}  e_a^{(r)} \right).
\nonumber
\end{eqnarray}
However, note that in all our approximations thus far we have only taken up to $(\alpha_r)^2$
to be non-negligible. We therefore need only consider, at most, the leading term in the sine, due to
it being squared in $r^2$. Hence, 
\begin{eqnarray}
\sum_a \sin^2 \left(\frac12 \sum_{r\neq 0} \alpha_r(t) e_a^{(r)} \right) &\approx&
\frac14 \sum_{r,r'\neq 0} \alpha_r \alpha_{r'} \sum_a e^{(r)}_a e^{(r')}_a \nonumber \\
&=& \frac14 \sum_{r\neq 0} \lambda_r (\alpha_r(t))^2 \nonumber
\end{eqnarray}
where {\it only} modes $r$ which are dynamical contribute to this.
For the steady state case where all modes are static no normal modes contribute:
\begin{eqnarray}
(r^{(\infty)})^2 \approx 1
\nonumber
\end{eqnarray}
and steady state is consistent with complete synchronisation.

For a single dynamical mode we have
\begin{eqnarray}
r_{1}^2 \approx 1 - \frac{1}{N^2} \lambda_1 \alpha_1^2(t)
\label{r1mode}
\end{eqnarray}
while for two modes the result is:
\begin{eqnarray}
r_{2}^2 \approx 1 - \frac{1}{N^2} (\lambda_1 \alpha_1^2(t) + \lambda_2 \alpha_2^2(t)).
\label{r2modes}
\end{eqnarray}
The generalisation to many dynamical modes is self-evident, though our analytic methods do not help
at this stage.

Drawing on our solutions to the various cases above we obtain 
either $r^2\approx 1$ or signals of periodic behaviours in the order parameter.
Therefore, if the system reaches a state where all but the 
two lowest Laplacian normal modes are suppressed a rich range of behaviours can occur
in the future evolution of the system:
complete synchronisation or partial synchronisation with the modes settling into limit cycles. 
Critical values of the coupling can be determined signalling
the boundary between oscillatory behaviour and complete synchronisation.
The approximation underlying these calculations is consistent with $\alpha \sim 1-2$
so that these effects should be detectable for either large graphs 
with very high algebraic connectivity, $\lambda_1>>1$, or small graphs with
algebraic connectivity of order unity.

\section{Simulation Studies} 
The aim of this section is not to show detailed simulation studies testing the
particular form of the critical coupling; our approximations do not warrant at this stage
such a study. Rather, the intent is to demonstrate cases of the transition between
incoherence, oscillatory behaviour and synchronisation consistent with the qualitative
picture we have developed. 

We use NetLogo 3.1.3 \cite{Netlogo}, a multi-agent modelling environment
developed by Uri Wilensky (a version 4.0.4 is currently available). 
In particular, this tool provides a number of models
for the study of complex networks as well as for synchronisation of fireflies.
Specifically, we use an extension of one of these models developed by Dekker \cite{Dekker07} 
which exploits the Kawachi \cite{Kaw04} process for generating a spectrum of well-studied complex
networks by variation of a single parameter $p$, a rewiring probability, starting with
a regular graph. For $p=0$ the graph is a ring, for $p=0.05-0.1$ small world networks
are produced, for $p=1-2$ the graph is random (not quite Erd{\"o}s-Renyi) and for
$p=5$ the graph is scale-free. This NetLogo implementation of the Kuramoto model 
sets up randomly selected initial phase angles and 
intrinsic frequencies drawn from a uniform
distribution in the range $0.02\pm w/2$. Thus the width $w$ of the distribution,
the coupling $\sigma$, the number of nodes $N$ and the type of graph (using $p$) are
selected prior to execution of 
a simulation. The position of phases around the circle are indicated in the simulation
graphic by a rotating colour scheme. In this model, ``correlations'' are
plotted as a function of time based on the original order parameter
of Kuramoto \cite{Kur84}, which can easily be shown
to display the same dynamical behaviours of our $r$ considered above.
The model is designed so that for any graph that can be dialled up a full range of behaviours,
from incoherence up to complete synchronisation, can be exhibited for a reasonable sweep
of the parameters of coupling and frequency distribution width.

We comment in more detail in the next section on the Laplacian spectral properties of various well-known
complex networks. Here it suffices to
say that for demonstrating the transition from incoherence to synchronisation we have selected
a scale-free graph ($p=5$) with $N=20$ nodes. The reason for this choice is to avoid high
suppression by inverse powers of $N$ of potentially oscillatory behaviour in $r$, see 
Eqs.(\ref{r1mode},\ref{r2modes}), while on the other hand avoiding a large build up of low-lying
Laplacian modes. We thus have a system close to being consistent with the conditions
of our approximations.

A number of screen captures from simulations are presented. We first show simulation for
the lowest possible coupling in the model settings, $\sigma=10^{-4}$, in Fig.{\ref{incoherent}. 
Though $r$ can occasionally reach large values there is no evidence of any order in the 
evolution of the fluctuations which is consistent with incoherence.

\begin{figure}[htb]
\includegraphics[width=85mm]{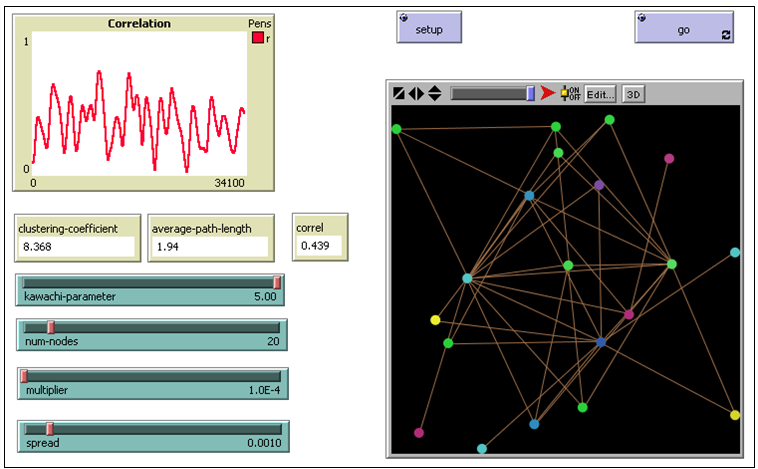}
\caption{NetLogo simulation of the Kuramoto model for a scale-free network of 20 nodes.
The coupling is set to $\sigma=10^{-4}$. 
The colours of the nodes in the network diagram visually represent the phase angles of
the oscillators and give a sense of how many oscillators are locked (the green nodes in this 
snapshot) versus how many are drifting (other colours). The order parameter ``correlation''
shows incoherent behaviour consistent with the poorly coupled network failing to synchronise.} 
\label{incoherent}
\end{figure}

We next show two cases at coupling of $\sigma=2\times 10^{-4}$
at which quite different periodic behaviours in the order
paramater can be seen. In the first instance, Fig.{\ref{period1}}, there is
evidently a single oscillating mode. This behaviour has been reproduced a number of times though
each case shows slightly different profiles according to the (small) number of nodes participating
in such modes. However, other
instances, for example shown in Fig.{\ref{period2}, 
exhibit two superposed periodic modes. Again,
a number of variations on this are observed for different random initial conditions
and intrinsic frequencies.

\begin{figure}[htb]
\includegraphics[width=85mm]{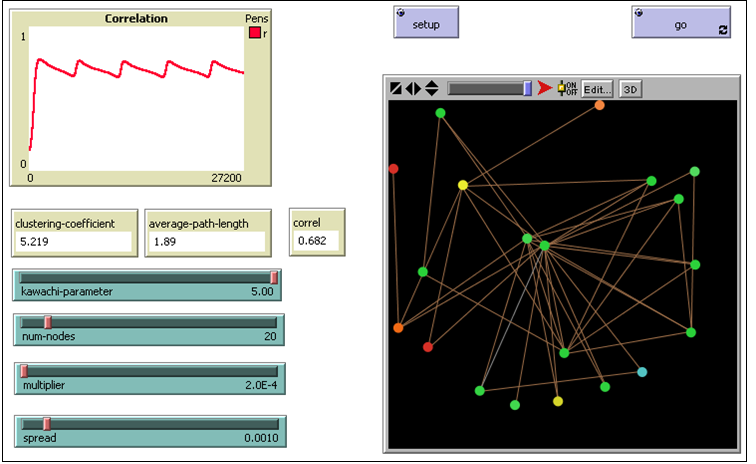}
\caption{NetLogo simulation of the Kuramoto model for a scale-free network of 20 nodes.
The coupling is set to $\sigma=2\times 10^{-4}$. The order parameter ``correlation''
shows periodic behaviour consistent with the limit cycle oscillation of a single mode.
From the colours of the nodes we can see that many nodes are locked (green) while several
(yellow and red) have not. This permits us to conclude that the limit cycle behaviour
is not a consequence of a single oscillating node but of collections (albeit in small numbers) of
nodes.} 
\label{period1}
\end{figure}

\begin{figure}[htb]
\includegraphics[width=85mm]{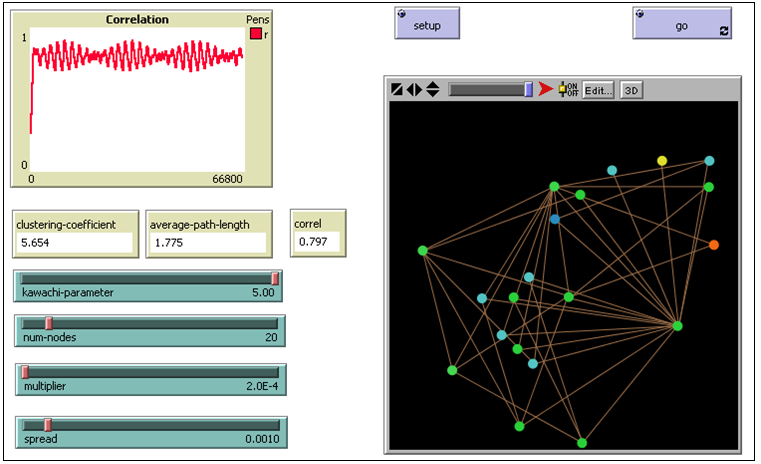}
\caption{NetLogo simulation of the Kuramoto model for a scale-free network of 20 nodes.
The coupling is set to $\sigma=2\times 10^{-4}$. The order parameter ``correlation''
shows periodic behaviour consistent with the limit cycle oscillation of two modes.} 
\label{period2}
\end{figure}

Finally, the coupling is
increased to $\sigma=5\times 10^{-4}$ whence rapid synchronisation occurs as seen in
Fig.{\ref{coherence}}. 

\begin{figure}[htb]
\includegraphics[width=85mm]{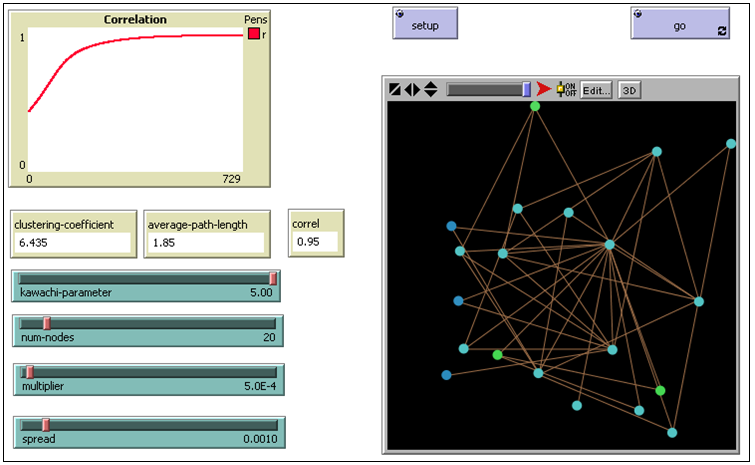}
\caption{NetLogo simulation of the Kuramoto model for a scale-free network of 20 nodes.
The coupling is set to $\sigma=5\times 10^{-4}$. The order parameter ``correlation''
shows rapid synchronisation. } 
\label{coherence}
\end{figure}

These behaviours are consistent with those identified analytically for the one and two
mode cases so that 
the periodicity is rather robust
though the precise form it takes can vary according to randomness within the network
and selection of intrinsic frequencies. Unfortunately, we are not able to extract
the network properties directly from NetLogo in order to check the low-lying Laplacian
spectral properties for these cases. 

\section{Discussion}

We have considered second order approximations about the
synchronised fixed point in the Kuramoto model on a general network. 
Motivated by insights from simulations that network sub-structures 
with smallest Laplacian eigenvalues
approach synchronisation later in time, we have 
assumed higher modes to have reached steady state and analysed the dynamical equations
for the remaining low-lying normal Laplacian modes. These equations resemble logistic
and Lotka-Volterra
equations and have yielded critical values for the coupling constant distinguishing
behaviours such as convergence to the fixed point and limit cycles. In particular,
this critical coupling is inversely proportional to the square of the
algebraic connectivity. For larger numbers of low-lying modes participating
in the dynamics more exotic dynamics can be exhibited, such as tori and chaos. In support of
these results for one and two modes
we have used NetLogo simulations to show evidence of periodic behaviour
as the coupling is increased. 
This shows that for dynamical processes on networks there
can be intermediate regimes of behaviour, between order (synchronisation) and chaos (incoherence),
exhibiting rich structured behaviour, an example of Edge of Chaos phenomena. 

At a broader level, these results show that for general
networks there is no single critical coupling for the onset of synchronisation
but many such coupling values indicating the regimes within which
substructures can stabilise into states of partial synchronisation.
We expect a similar hierarchy of couplings to be reflected in the nonperturbative
regime, where
the bulk of the normal modes have their dynamics ``extinguished''. 
Our lack of analytical handle on the
mechanism at play in this regime means we cannot yet compare with the case where this model
all began, the complete graph: the high degeneracy of Laplacian eigenvalues here (infinite degeneracy
for Kuramoto's $N\rightarrow\infty$ case) means that
the normal modes' capacity to reach steady state lies entirely in this nonperturbative regime.

However, we can speculate to some degree on the role of Laplacian modes here. We notice
that, whereas for regular graphs the Laplacian spectra are quite dispersed with multiple
gaps between groups of (often degenerate) eigenvalues, 
for all known complex networks there occurs a clear accumulation of modes
about some eigenvalue \cite{JamVanM}: a ``bulk'' of modes is present in the spectrum. 
Erd{\"o}s-Renyi random graphs generated by a low probability of creating a 
random link between any two nodes
show this bulk at small eigenvalues while for
large probability this bulk shifts up to higher values. Watts-Strogatz
networks with small probability for rewiring a lattice graph 
have the bulk in a narrow curve around the mean
nodal degree with smaller peaks reflecting the original spectrum for the lattice 
while for larger probability the main peak broadens. Finally, scale-free graphs have the main
bulk occuring narrowly around eigenvalues of order unity with a few eigenvalues less than one. 
This bulk in the Laplacian spectrum
is the common feature of any graph which, according to numerical studies,
readily demonstrates synchronisation at finite values of the coupling.

A nonperturbative mechanism is therefore suggested: close-lying
eigenmodes somehow mutually interfere with each other such as to cause them to de-excite to
their steady states; when the
majority of eigenvalues are closely spaced this de-excitation results in a large number of modes
reaching steady state values. The threshold for this de-excitation is given by a critical coupling
value. For the complete graph, the complete degeneracy of modes means this de-excitation across
modes is nearly instantaneous. For other complex graphs this de-excitation takes some time to
propagate, leaving some residual number of dynamical low-lying modes whose suppression has
been the subject of this paper.

This insight suggests new approximations of the dynamical equations which may provide
a handle on this nonperturbative mechanism. Our future work will seek to provide more substantial 
support to this hypothesis.

\section*{Acknowledgements}
The author is indebted to fruitful discussions with Brian Hanlon and Richard Taylor
as well critical feedback and assistance with the NetLogo simulator from Anthony Dekker.


\begin{thebibliography}{9}
\bibitem{Kall08}
A. Kalloniatis, A new paradigm for modelling networked dynamical
systems, 13th International Command and Control Research and
Technology Symposium, Seattle USA, (2008)
\bibitem{Wie58}
N. Wiener, Nonlinear Problems in Random Theory, MIT Press, Cambridge
(1958)
\bibitem{Win67}
A.T. Winfree, J.Theoret.Biol, 16, 15 (1967).
\bibitem{Kur84}
Y. Kuramoto, Chemical Oscillations, Waves and Turbulence, Springer,
Berlin (1984).
\bibitem{Stro00}
S.H. Strogatz, Physica D 143, 1 (2000).
\bibitem{Ace05}
J.A. Acebron, L.L. Bonilla, C.J. Perez Vicente, F. Ritort, R. Spigler, Rev.Mod.Phys.77, 137 (2005)
\bibitem{Ich04}
T. Ichinomiya, Phys.Rev.E 70, 026116 (2004)
\bibitem{Gomez07}
J. Gomez-Gardenes, Y. Moreno, A. Arenas, Phys.Rev.E, 066106 (2007)
\bibitem{AlbBar02}
R. Albert, A.-L. Barab{\'a}si, Rev.Mod.Phys. 74, 47 (2002)
\bibitem{WattsStrog98}
D.J. Watts, S.H. Strogatz, Nature, 393, 440 (1998)
\bibitem{Hong02}
H. Hong, M.Y. Choi, B.J. Kim, Phys.Rev.E 65, 047104 (2002)
\bibitem{Aren06}
A. Arenas, A. Diaz-Guilera, C.J. Perez-Vicente,
Phys. Rev. Lett. 96, 114102 (2006)
\bibitem{Chung97}
F.R.K. Chung, Spectral Graph Theory, CBMS, Regional Conference Series
in Mathematics, No. 92, American Mathematical Society, Providence
(1997)
\bibitem{Boll98}
B. Bollob{\'a}s, Modern Graph Theory, Graduate Texts in Mathematics, Springer, USA, 1998 
\bibitem{Dor08}
S.N. Dorogovstev, A.V. Goltsev, Rev.Mod.Phys. 80, 1275 (2008)
\bibitem{Mir05}
R.E. Mirollo, S.H. Strogatz, Physica D, 205, 249 (2005)
\bibitem{Fied73}
M. Fiedler, Czech. Math. J., 23, 298 (1973).
\bibitem{Pec98}
L.M. Pecora, T.L. Carroll, Phys.Rev.Lett. 80, 2109 (1998)
\bibitem{various}
A.E. Motter, C. Zhou, J. Kurths, Phys.Rev.E 71, 016116 (2005);
T. Nishikawa, A.E. Motter, Phys.Rev.E 73, 065106 (2006);
B. Gong, L. Yang, K. Yang, Phys.Rev.E, 037101 (2005);
Z.Li, G. Chen, IEEE Transactions on Circuits and Systems, 53, 28 (2006)
\bibitem{Ding01}
C.H.Q. Ding, X. He, H. Zha, Proceedings of the seventh ACM SIGKDD
international conference on 
Knowledge discovery and data mining, p.275 (2001)
\bibitem{Jad04}
A. Jadbabaie, N. Motee, M. Barahona, Proceedings of the American
Control Conference, Vol. 5., 4296 (2004).
\bibitem{Kap95} 
D. Kaplan, L. Glass, Understanding Nonlinear Dynamics, Springer, NY (1995)
\bibitem{Smal76} 
S. Smale, J.Math.Biol. 3,5 (1976)
\bibitem{SWA05} 
J.C.Sprott, J.C. Wildenberg, Y. Azizi, Chaos, Solitons and Fractals, 26, 1035 (2005)
\bibitem{Netlogo}
See http://ccl.northwestern.edu/netlogo.
\bibitem{Dekker07} 
A. Dekker, Journal of Artificial Societies and Social Simulation,
10 (4) 6; http://jasss.soc.surrey.ac.uk/10/4/6.html; the model can be downloaded from
http://jasss.soc.surrey.ac.uk/10/4/6/resources/
NewKawachi.nlogo
\bibitem{Kaw04} 
Y. Kawachi, K. Murata, S. Yoshi, Y. Kakazu, 
Proc. 7th Asia-Pacific Conf. on Complex Systems, Australia, pp. 247-255, 2004
\bibitem{JamVanM} 
A. Jamakovic, P. van Mieghem, European Conference on Complex Systems, (2006)
\end{thebibliography}
\end{document}